\documentclass[10pt, twocolumn, twoside]{IEEEtran}

\usepackage{cite}      

\usepackage{graphicx}  
\usepackage{amssymb}
\usepackage{verbatim}
\usepackage{graphicx}
\usepackage{amsmath}  
\usepackage{bm} 
\usepackage{float}

\newcommand{\Mp}{\boldsymbol{\phi}}

\floatstyle{ruled}
\newfloat{algorithm}{htb}{lop}
\floatname{algorithm}{Algorithm}

\begin{document}
\title{Compressive Sensing Using Low Density Frames}
\author{Mehmet~Ak\c{c}akaya,
				~Jinsoo~Park~and~Vahid~Tarokh
\thanks{M. Ak\c{c}akaya, J. Park and V. Tarokh are with the School of Engineering and Applied Sciences, Harvard University, Cambridge, MA, 02138. (e-mails: \{akcakaya, vahid\}@seas.harvard.edu, park10@fas.harvard.edu)}}
\markboth{DRAFT}{DRAFT}
\maketitle

\begin{abstract}
We consider the compressive sensing of a sparse or compressible signal ${\bf x} \in {\mathbb R}^M$. We explicitly construct a class of measurement matrices, referred to as the low density frames, and develop decoding algorithms that produce an accurate estimate $\hat{\bf x}$ even in the presence of additive noise. Low density frames are sparse matrices and have small storage requirements. Our decoding algorithms for these frames have $O(M)$ complexity. Simulation results are provided, demonstrating that our approach significantly outperforms state-of-the-art recovery algorithms for numerous cases of interest. In particular, for Gaussian sparse signals and Gaussian noise, we are within 2 dB range of the theoretical lower bound in most cases.
\end{abstract}

\IEEEpeerreviewmaketitle

\section{Introduction}

\PARstart{L}{et} ${\bf x} = (x_1, \dots, x_M) \in {\mathbb R}^M$ with $||{\bf x}||_0 = |\{x_i: x_i \neq 0\}| = L$. ${\bf x}$ is said to be sparse if $L << M$. Consider the equation 
\begin{equation} \label{noisy_cs_eq}
{\bf y = Ax + n},
\end{equation}
where ${\bf A}$ is an $N \times M$ measurement matrix and ${\bf n} \in {\mathbb R}^N$ is a noise vector. When $N < M$, ${\bf y}$ is called the compressively sensed version of ${\bf x}$ with measurement matrix ${\bf A}$. In this paper, we are interested in coming up with a good estimate $\hat{\bf x}$ for a sparse vector ${\bf x}$ from the observed vector ${\bf y}$ and the measurement matrix ${\bf A}$.

We refer to the case ${\bf n = 0}$ as \emph{noiseless compressive sensing}. This is the only case when ${\bf x}$ can be perfectly recovered. In particular, 
it can be shown \cite{Candes-Tao, Needell3} that if ${\bf A}$ has the property that every of its $N$ columns are linearly independent, then a decoder can recover ${\bf x}$ uniquely from $N = 2L$ samples by solving the $\ell_0$ minimization problem
\begin{equation}  \label{ell0_eq}
\min ||{\bf x}||_0 \:\:\: \textrm{ s. t. } \:\:\: {\bf y = Ax}.
\end{equation}

However, solving this $\ell_0$ minimization problem for general ${\bf A}$ is NP-hard \cite{Tropp}. An alternative solution method proposed in the literature is the $\ell_1$ regularization approach, where 
\begin{equation} \label{ell1_eq}
\min ||{\bf x}||_1 \:\:\: \textrm{ s. t. } \:\:\: {\bf y = Ax},
\end{equation}
is solved instead. Criteria has been established in the literature as to when the solution of (\ref{ell1_eq}) coincides with that of (\ref{ell0_eq}) in the noiseless case for various classes of measurement matrices \cite{Candes-Tao, Donoho2}. In an important contribution, for ${\bf A}$ belonging to the classes of Gaussian and partial Fourier ensembles, Cand\`es and Tao showed \cite{Candes-Tao} that this recovery problem can be solved for $L = O(M)$ with $N=O(L)$ as long as the observations are not contaminated with (additive) noise. 

It can be shown that there is a relationship between the solution to problem (\ref{noisy_cs_eq}) and minimum Hamming distance problem in coding theory \cite{AkTar, AkTar-isit, Baron1}. This approach was further exploited in \cite{Xu}. Using this connection, we constructed ensembles of measurement matrices\footnote{We use the terms ``frame'' and ``measurement matrix'' interchangeably throughout the rest of the paper.} and associated decoding algorithms that solves the $\ell_0$ minimization problem with complexity $O(MN)$ for $L = O(M)$ with $N=O(L)$ in the noiseless case \cite{AkTar, AkTar-isit}.

For problem (\ref{noisy_cs_eq}) with non-zero ${\bf n}$, referred to as \emph{noisy compressive sensing}, the $\ell_1$ regularization approach of (\ref{ell1_eq}) can also be applied. For a measurement matrix ${\bf A}$ that satisfies a property called the restricted isometry principle (RIP), the quadratic program $$\min ||{\bf x}||_1 \:\:\: \textrm{ s. t. } \:\:\: ||{\bf Ax - y}||_2 \leq \epsilon,$$
can be solved for a parameter $\epsilon$ related to $||{\bf n}||_2$, and an estimate $\hat{\bf x}_{\textrm{QP}}$ can be obtained such that $||\hat{\bf x}_{\textrm{QP}} - {\bf x}||_2 \leq C_1\: \epsilon,$ where $C_1$ is a constant that depends on ${\bf A}$ \cite{Candes-Romberg-Tao}. If ${\bf n} \sim {\cal N}(0, \sigma^2 {\bf I}_N)$, another approach is the Dantzig Selector
$$\min ||{\bf x}||_1 \:\:\: \textrm{ s. t. } \:\:\: ||{\bf A^{*}(Ax - y)}||_{\infty} \leq \gamma,$$
where $\gamma$ is a function of $\sigma$ and $M$. This gives an estimate $\hat{\bf x}_{\textrm{DS}}$ such that ${\mathbb E}_{\bf n}||\hat{\bf x}_{\textrm{DS}} - {\bf x}||_2^2 \leq C_2 (\log M) \sum_i \min(x_i^2, \sigma^2),$ where $C_2$ is a constant that depends on ${\bf A}$ \cite{Dantzig}. 
Both these methods may not be easily implemented in real time with the limitations of today's hardware. 
To improve the running time of $\ell_1$ methods, some authors have investigated using sparse matrices for ${\bf A}$  \cite{Berinde2}. Using the expansion properties of the graphs represented by such matrices, it was shown in \cite{Berinde2} that it is possible to obtain an estimate $\hat{\bf x}_{\textrm{E}}$ such that $||\hat{\bf x}_{\textrm{E}} - {\bf x}||_1 \leq C_3 ||{\bf n}||_1$, where $C_3$ is a constant that depends on ${\bf A}$. 

Another strand of work studies recovery algorithms based on the matching pursuit algorithm \cite{Gilbert}. Recently, variants of this algorithm, e.g. Subspace Pursuit \cite{Dai} and CoSaMP \cite{Needell3}, have been proposed. Both algorithms provably work for measurement matrices satisfying RIP, and guarantee perfect reconstruction in the noiseless setting for $N = O(L \log(M/L))$ as the $\ell_1$ recovery methods do. For the noisy problem, they also offer similar guarantees to $\ell_1$ methods. 
These algorithms have complexity $O({\cal L} \log R)$, where ${\cal L}$ is the complexity of matrix-vector multiplication ($O(MN)$ for Gaussian matrices, $O(N\log N)$ for partial Fourier ensembles) and $R$ is a precision parameter bounded above by $||{\bf x}||_2$ (which is $O(N)$ for a fixed signal-to-noise ratio). In \cite{Berinde3}, Sparse Matching Pursuit (SMP) was proposed for sparse ${\bf A}$ and this algorithm has $O(M \log(M/L))$ complexity .

Yet another direction in compressive sensing is the use of the Bayesian approach. In \cite{Carin}, the idea of relevance vector machine (RVM) \cite{Tipping} has been applied to compressive sensing. Although simulation results indicate that the algorithm has good performance, it has complexity $O(MN^2)$.

In this paper, we study the construction of measurement matrices that can be stored and manipulated efficiently in high dimensions, and fast decoding algorithms that generate estimates with small $\ell_2$ distortion. The ensemble of measurement matrices are a generalization of LDPC codes and we refer to them as \emph{low density frames} (LDFs). For our decoding algorithms, we combine various ideas from coding theory, statistical learning theory and theory of estimation. Simulation results are provided indicating an excellent distortion performance at high levels of sparsity and for high levels of noise.

The outline of this paper is given next. In Section \ref{sec:LDF}, we introduce low density frames and study their basic properties. In Section \ref{sec:suprem}, we introduce various concepts used in algorithms and describe the decoding algorithms. In Section \ref{sec:sims}, we provide extensive simulation results for a number of different testing criteria. Finally in Section \ref{sec:conc}, we make our conclusions and provide directions for future research.


\section{Low Density Frames} \label{sec:LDF}

Let ${\mathcal{F}} = \{ \Mp_1, \Mp_2, \cdots, \Mp_M \}$ be a frame consisting of
$M \ge N $ non-zero vectors which span ${\mathbb R}^N$.
Let $\Mp_i = (\phi_{1,i}, \cdots, \phi_{N,i})$  for $i=1,2, \cdots, M$.
This frame could be represented in matrix form as an $N \times M$ matrix
\begin{eqnarray}
{\bf F} = \left( \begin{array}{cccc}
\phi_{1,1} & \phi_{1,2} & \cdots  & \phi_{1,M} \\ 
\phi_{2,1} & \phi_{2,2} & \cdots  &\phi_{2,M}  \\
\vdots  & \ddots & \ddots & \vdots \\
\phi_{N,1} & \phi_{N,2} & \cdots & \phi_{N,M} 
\end{array} \right).
\end{eqnarray} 

A {\it low density frame} (LDF) ${\cal F}$ is defined by a matrix ${\bf F}$ where the vast majority of elements of each column and each row of ${\bf F}$ are zeroes. 
Formally, we define a $(d_v, d_c)$-{\it regular} LDF as a matrix ${\bf F}$ that has $d_c$ non-zero elements in each row and $d_v$ non-zero elements in each column. Clearly $M d_v = N d_c$. We also note that the redundancy of the frame is $r = M/N = d_c/d_v$. We will restrict ourselves to {\it binary} regular LDFs, where the non-zero elements of ${\bf F}$ are ones. 

The density $\rho$ of a frame ${\bf F}$ is the ratio of the number of non-zero entries of ${\bf F}$ to the dimension of ${\bf F}$. In this paper, we consider regular LDFs for which $\rho = (M d_v)/(MN) = d_v / N << 1$. 

As with LDPC codes, it is natural to represent LDFs using bipartite graphs. Furthermore, there is a well-established literature on inference in graphical models. Some of these methods can be used as a basis for recovery algorithms in the context of compressive sensing. To this end, we next summarize two important ideas from graphical models, namely factor graphs and the sum-product algorithm, and show how LDFs can be viewed as factor graphs.

\subsection{Factor Graphs}

Factor graphs are used to represent factorizations of functions of several variables \cite{Bishop, FactorGraphs}. Let $f({\bf w})$ be a function of several variables that can be factored as 
\begin{equation} \label{fac1}
f({\bf w}) = \prod_{s} f_s ({\bf w}_s).
\end{equation}
In this factorization each \emph{factor} $f_s$ is only a factor of ${\bf w}_s$, the subset of \emph{variable nodes} ${\bf w}$. 

A \emph{factor graph} depicting (\ref{fac1}) consists of variable nodes represented by circles, factor nodes represented by bold squares, and undirected edges connecting each factor node to all the variable nodes involved in that factor. 

\begin{figure}
\centering
\includegraphics[width=1.2in]{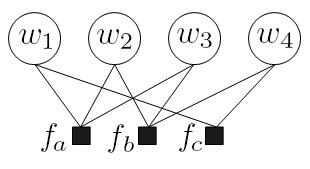}
\caption{Factor graph representing the function in Equation \ref{basfac}.}
\label{factor_basic}
\vspace{-.15cm}
\end{figure}

As an example, the factor graph representing 
\begin{equation} \label{basfac}
f({\bf w}) = f_a (w_1, w_2, w_3) f_b(w_2, w_3, w_4) f_c(w_1, w_4) \end{equation} is depicted in Figure \ref{factor_basic}.


\subsection{Sum-Product Algorithm}

The natural inference algorithm for factor graphs is the sum-product algorithm \cite{Bishop, Gallager}. This algorithm is an exact interference algorithm for tree-structured graphs (i.e. graphs with no cycles), and is usually described over discrete alphabets. However, the ideas also apply to continuous random variables with the sum being replaced by integration. In doing so, the computational cost of implementation increases and this issue will be addressed later. 

Suppose the goal is to find the marginal density $p(w)$ for a particular variable $w$. In particular, we have $$p(w) = \sum_{{\bf w} \setminus w} p({\bf w}).$$ 

One treats $w$ to be the root node of a tree, and looks at the subtrees connected to $w$ via factor nodes. Using this approach, the joint distribution can be written as 
\begin{equation} \label{factor1} 
p({\bf w}) = \prod_{s \in ne(w)} F_s(w, W_s),
\end{equation}
where $ne(w)$ is the neighborhood of $w$, i.e. the set of factor nodes that are connected to $w$, and $W_s$ is the set of variable nodes in the subtree connected to the factor node $f_s$ in $ne(w)$ \cite{Bishop}. $F_s(w, W_s)$ represents the product of the factors in the subtree associated with $f_s$.
Interchanging the summation and the products yields
 $$p(w) = \prod_{s \in ne(w)} \mu_{f_s \to w} (w),$$
where $\mu_{f_s \to w} (w)$ is the message sent from factor node $f_s$ to variable node $w$. One can show \cite{Bishop} that 
$$ \mu_{f_s \to w} (w) = \sum_{{\bf w}_s \setminus w} f_s(w, {\bf w}_s) \prod_{m \in ne(f_s) \setminus w} \mu_{w_m \to f_s} (w_m),$$
where ${\bf w}_s$ are all variable nodes connected to the factor node $f_s$, including $w$, and $ne(f_s)$ are the set of variable nodes connected to the factor node $f_s$. One can also show \cite{Bishop}
$$\mu_{w_m \to f_s} (w_m) = \prod_{l \in ne(w_m)\setminus f_s} \mu_{f_l \to w_m} (w_m).$$

Thus there are two types of messages, one type going from factor nodes to variable nodes denoted $\mu_{f \to w}$ and the other going from variable nodes to factor nodes denoted $\mu_{w \to f}$. The message propagation starts from the leaves of the factor graph. A leaf variable node sends an identity message $\mu_{w \to f}(w) = 1$ to its parent, whereas a leaf factor node $f$ sends $\mu_{f \to w}(w) = f(w)$, a description of $f$ to its parent. These expressions for messages give an efficient way of calculating the marginal probability distribution. We note that in writing out the factorization in (\ref{factor1}), it is essential that the graph has a tree structure so that the factors in the joint probability distribution $p({\bf w})$ can be partitioned into groups, each of which is associated with a single factor node in $ne(w)$.

The algorithm is easily modified to calculate the marginal for every variable node in the graph \cite{Bishop}. This modification results in only twice as many calculations as calculating a single marginal. A more interesting case is when there are observed variables in the graph, ${\bf v}$. In this case the sum-product algorithm could be used to calculate posterior marginals $p(w_i | {\bf v} = \hat{\bf v})$.



\subsection{Graphical Representation of Low Density Frames} 

The main connection between the $\ell_0$ minimization problem and coding theory involves the description of the underlying code \cite{AkTar}, ${\cal V}$ of ${\bf F}$, where 
$${\cal V} = \{{\bf d} \in {\mathbb R}^M: {\bf Fd = 0}\}.$$
One can view ${\cal V}$ as the set of vectors whose product with each row of ${\bf F}$ ``checks'' to $0$. In the works of Tanner, it was noted that this relationship between the checks and the codewords of a code can be represented by a bipartite graph \cite{Tanner}. This bipartite graph consists of two disjoint sets of vertices, $V$ and $C$, where $V$ contains the variable nodes and $C$ contains the factor nodes representing checks that codewords need to satisfy. Thus we have $|V| = M$ and $|C| = N$. Furthermore node $j$ in $V$ will be connected to node $i$ in $C$ if and only if the $(i,j)^{\textrm{th}}$ element of ${\bf F}$ is non-zero. Thus the number of edges of the graph is equal to the number of non-zero elements in the measurement matrix ${\bf F}$. For an LDF, this leads to a sparse bipartite graph. 

\begin{figure}
\centering
\includegraphics[width=3.2in]{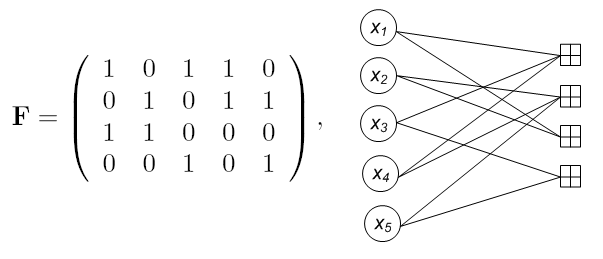}
\caption{A frame ${\bf F}$ and its graphical representation.}
\label{basic_ldpc}
\vspace{-.15cm}
\end{figure}

A simple example of this graphical representation is depicted in Figure \ref{basic_ldpc}. For representation of LDFs, it is convenient to use a factor node, depicted by a $\boxplus$, called a parity check node. This node has the property that the variable nodes connected to it should sum to zero. We also note that for the purposes of decoding, it is more convenient to use syndromes \cite{Mackay-book} that represent the measurement vector, ${\bf r}$. This is done by connecting a variable node representing the $j^{\textrm{th}}$ component of ${\bf r}$ to the $j^{\textrm{th}}$ check node. In this case, the parity check node has the property that the variable nodes connected to it sum to $r_j$. Thus the graph now represents the set $\{{\bf d} \in {\mathbb R}^M: {\bf Fd = r}\}$ which is a coset of the underlying code of the frame.

It is important to note that the graph representing an LDF will have cycles. Without the tree structure, the sum-product algorithm will only be an approximate inference algorithm. However, it has been empirically shown that for sparse graphs this approximate algorithm works very well \cite{Mackay, Urbanke, Wiberg}.


\section{Sum Product Algorithm with Expectation Maximization} \label{sec:suprem}

It is well-known in coding theory literature, that the standard decoding algorithm for codes on graphs is the sum-product algorithm (SPA) \cite{Gallager, FactorGraphs, Mackay}. Given a set of observations, this algorithm can be used to approximate the posterior marginal distributions. In fact, when there is no noise, variants of this algorithm \cite{Sipser} has been successfully adapted to compressive sensing \cite{Baron1, Xu}. However, when we are interested in the practical case of noisy observations, these algorithms no longer can be applied in a straightforward manner. Some authors have tried to circumvent this difficulty by using a two-point Gaussian mixture approach \cite{Baron2}, however the complexity of this algorithm may grow quickly as the number of Gaussian components in the mixtures could grow exponentially, unless some approximation is applied. However, using these approximations degrades the performance of the LDF approach. 

In this paper, we consider Gaussian Scale Mixture (GSM) priors with Jeffreys' non-informative hyperprior to obtain an algorithm that provides estimates for the noisy compressive sensing problem $${\bf r= Fx + n},$$ as well as the noiseless problem. Throughout the paper we assume that $${\bf n} \sim {\cal N}(0, \sigma^2 {\bf I}_N).$$ However simulation results (not included in this paper) indicate that the algorithms still work well even for non-Gaussian noise. We define the signal-to-noise ratio (SNR) as $$\textrm{SNR} = 10 \log_{10} \frac{||{\bf Fx}||^2}{{\mathbb E}_{\bf n}||{\bf n}||^2} = 10 \log_{10} \frac{||{\bf Fx}||^2}{\sigma^2 N}.$$

\subsection{Gaussian Scale Mixtures} \label{sec:GSM}

The main difficulty in using the sum-product algorithm (SPA) in compressive sensing setting is that the variables of interest are continuous. Nonetheless SPA can be employed naturally when the underlying continous random variables are Gaussian  \cite{Weiss}. Since any Gaussian pdf ${\cal N}(x |a, A)$ can be determined by its mean $a$ and variance $A$, these constitute the messages in this setting. At the variable nodes, the product of Gaussian probability density functions (pdf) will result in a (scaled) Gaussian pdf, and at the check nodes, the convolution of Gaussian pdfs will also result in a Gaussian pdf. i.e. $${\cal N}(x | a_1, A_1) \ast {\cal N}(x | a_2, A_2) \propto {\cal N}(x | a_1+a_2 , A_1+A_2),$$ and $${\cal N}(x | a_1, A_1) \cdot {\cal N}(x | a_2, A_2) \propto {\cal N}(x | b, B),$$
where $\propto$ denotes normalization up to a constant, and $$ B = (A_1^{-1} + A_2^{-1})^{-1}, $$ $$b = B (A_1^{-1} a_1 + A_2^{-1} a_2).$$
We note that all the underlying operations for SPA preserve the Gaussian structure. 

It is well-known that the Gaussian pdf is not ``sparsity-enhancing''. Thus some authors propose the use of the Laplacian prior
\begin{equation} \label{Laplacian}
p({\bf x}) = \prod_i p_{x_i}(x_i) = \prod_i \frac{\lambda}{2} \exp(- \lambda |x_i|).
\end{equation}

Clearly with this prior and for Gaussian noise ${\bf n}$
\begin{equation} \label{laplace-lasso}
p({\bf x}|{\bf y}) \propto p({\bf y}|{\bf x}) p({\bf x}) \propto \exp\big(- ||{\bf y - Ax}||_2^2 - \lambda' ||{\bf x}||_1\big),
\end{equation}
and maximization of $p({\bf x}|{\bf y})$ is equivalent to minimizing $$||{\bf y - Ax}||_2^2 + \lambda' ||{\bf x}||_1,$$ which is the objective function for the LASSO algorithm \cite{Tropp2,Wainwright2}. However, a straightforward implementation of this algorithm may not be computationally feasible.

In this paper, we consider the family of Gaussian Scale Mixtures (GSM) densities \cite{Andrews}, given by
$$ x = \sqrt{\beta} u, $$ 
where $u$ is a zero-mean Gaussian and $\sqrt{\beta}$ is a positive scalar random variable. Hence
$$p_{x | \beta} (x|\beta) \sim {\cal N}(x | 0, \beta), $$ 
and $$p_x(x) = \int_0^{\infty} p_{x | \beta} (x|\beta) p_{\beta} (\beta) d\beta. $$

This family of densities are symmetric, zero-mean and have heavier tails than a Gaussian, 
and have been successfully used in image processing \cite{Dias, Nowak2, Portilla}, and learning theory \cite{Tipping}.

In order to completely specify our model, we need to choose a pdf for $p_{\beta}(\beta)$. In this paper, we use 
$$p_{\beta}(\beta) \propto \sqrt{\det(I(\beta))}, \quad I(\beta) = {\mathbb E}\Bigg(-\frac{\partial^2 \log p_{x|\beta}(x|\beta)}{\partial \beta^2} \bigg| \beta \Bigg)$$
where $I(\beta)$ is the Fisher information matrix. This is referred to as the Jeffreys' prior, which can be shown to be a scalar invariant prior suitable for sparse estimation \cite{Robert}. 
In our model, the prior is given by $$p_{\beta_i} (\beta_i) = \frac{1}{\beta_i},$$ which has no parameters to optimize. We note that this is an improper density, i.e. it cannot be normalized. In Bayesian statistics, these kind of improper priors are used frequently, since only the relative weight of the prior determines the a-posteriori density \cite{Robert}. This density also has a singularity at the origin. This fact is usually ignored as long as it does not create computational problems \cite{Portilla}. As an alternative one might set the prior to $0$ in a small interval $\beta \in [0,\beta_{min})$. We also note that with this choice for $p_{\beta_i} (\beta_i)$, $p_{x_i} (x_i) \propto 1 / |x_i|$, which is a very heavy-tailed density. 

\begin{figure}
\centering
\includegraphics[width=3.5in]{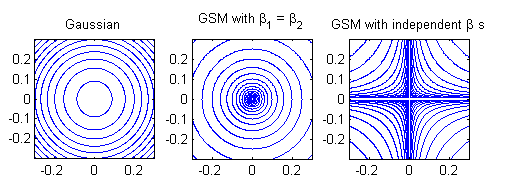}
\caption{Contour plots for a Gaussian distribution (left), a GSM with $\beta_1 = \beta_2$ 
distributed according to Jeffreys' prior (middle), a GSM with $\beta_1$ and $\beta_2$ i.i.d. with Jeffreys' prior (right).}
\label{indep_betas}
\vspace{-.2cm}
\end{figure}

To enhance sparsity in each coordinate, it is important to have independent $\beta_i$ for all $i$ \cite{Tipping2}. As depicted in the middle subplot of Figure \ref{indep_betas}, compared to a Gaussian distribution, a GSM with $\beta_i$ distributed according to Jeffreys' prior has a much sharper peak at the origin even when $\beta_1 = \beta_2$. However, the subplot on the right demonstrates that if the $\beta_i$s are indeed independent, the GSM will be highly concentrated not only around the origin, but along the coordinate axes as well, which is a desired property if we have no further information about the locations of the sparse coefficients of ${\bf x}$. In our model, we will assume that $$p({\bf x}, {\bm \beta}) = \prod_{i=1}^M p(x_i | \beta_i) \prod_{i=1}^M p(\beta_i)$$ in order to enhance sparsity in all coordinates. 
This independence assumption is natural and commonly used in the literature \cite{Nowak2, Tropp2, Wainwright2}.

\subsection{Expectation Maximization}

The expectation maximization (EM) algorithm is a method for finding maximum-likelihood (ML) estimates of parameters in a model with observed and hidden variables \cite{Moon}. Let ${\bf y}$ be the observed data and let ${\bf z}$ be the hidden data. Let the probability density function of $({\bf y,z})$ be $f({\bf y,z}|{\bm \theta})$, parametrized by the vector ${\bm \theta}$. The EM algorithm iteratively improves on an initial estimate ${\bm \theta}^{(0)}$ using a two-step procedure. In the expectation step (E-step), we calculate 
$$Q({\bm \theta} | {\bm \theta}^{(k)}) = {\mathbb E}_{\bf z} \Big(\log f({\bf y,z}|{\bm \theta}) | {\bf y}, {\bm \theta}^{(k)} \Big)$$
given an estimate ${\bm \theta}^{(k)}$ from the previous iteration. It is important to distinguish the two arguments of the $Q$ function  are different. The second argument is the conditioning argument for the expectation and is fixed during the E-step. In the second step, called the maximization step or M-step, a new estimate 
$${\bm \theta}^{(k+1)} = \arg \max_{\bm \theta} Q({\bm \theta} | {\bm \theta}^{(k)})$$
is calculated.

It can be shown that the estimates monotonically increases the likelihood with respect to the observed data ${\bf y}$ \cite{Moon}, $$ f({\bf y} | {\bm \theta}^{(k+1)}) \geq f({\bf y} | {\bm \theta}^{(k)}).$$

When ${\bm \theta}$ is itself a random variable, the M-step maximizes $\big(Q({\bm \theta} | {\bm \theta}^{(k)}) + \log f({\bm \theta})\big)$, and the EM algorithm can be used to find a maximum a-posteriori (MAP) estimate of ${\bm \theta}$ \cite{Krishnan}.

\subsection{SuPrEM Algorithm I}

\begin{figure}
\centering
\vspace{-.2cm}
\includegraphics[width=3.5in]{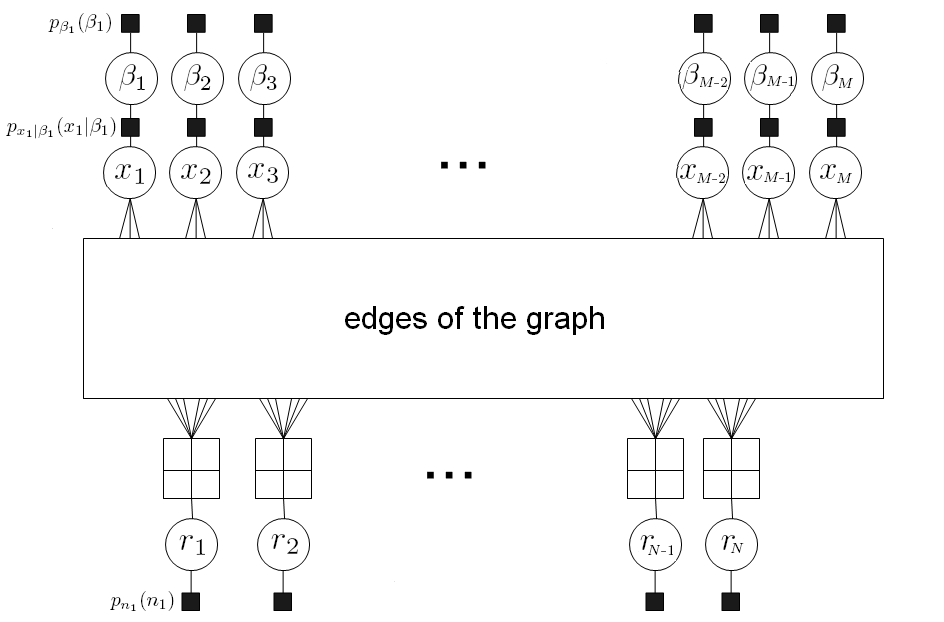}
\caption{The factor graph for a (3,6)-regular LDF with the appropriate hyperpriors.}
\label{suprem_FG}
\vspace{-.2cm}
\end{figure}

The factor graph for decoding purposes is depicted in Figure \ref{suprem_FG}. Here, ${\bf r}$ is the vector of observed variables, ${\bf x}$ is the vector of hidden variables and ${\bm \beta}$ is the vector of parameters. 
We next propose the \emph{Sum Product with Expectation Maximization (SuPrEM)} Algorithm I. At every iteration $t$, this algorithm uses a combination of the Sum-Product Algorithm (SPA) and EM algorithm to generate estimates for the hyperpriors $\{\beta_k^{(t)}\}$, as well as a point estimate $\{\hat{x}_k^{(t)} \}$. In the EM stage of the algorithm, $Q({\bm \beta} | {\bm \beta}^{(t)})$ for the \textbf{E-step} is given by 

\begin{align}
	Q({\bm \beta} &| {\bm \beta}^{(t)}) = {\mathbb E}_{\bf x} \bigg( \log p({\bf x, y}, {\bm \beta}) \big| {\bf y}, {\bm \beta}^{(t)} \bigg) \nonumber\\
		&= {\mathbb E}_{\bf x} \bigg( \log \big( p({\bf y}|{\bf x}) p({\bf x}|{\bm \beta}) p({\bm \beta}) \big) \big| {\bf y}, {\bm \beta}^{(t)} \bigg) \nonumber\\
		&= {\mathbb E}_{\bf x} \bigg( \log p({\bf y}|{\bf x})\big| {\bf y}, {\bm \beta}^{(t)} \bigg) + {\mathbb E}_{\bf x} \bigg( \log p({\bf x}, {\bm \beta}) \big| {\bf y}, {\bm \beta}^{(t)} \bigg) \nonumber\\
		&= C_1 + \sum_{i=1}^M {\mathbb E}_{\bf x} \bigg( \log p(x_i ,\beta_i) \big| {\bf y}, {\bm \beta}^{(t)} \bigg) \nonumber\\
		&= C_1 + \sum_{i=1}^M {\mathbb E}_{x_i} \bigg( \log p(x_i ,\beta_i) \big| {\bf y}, {\bm \beta}^{(t)} \bigg),
\end{align}
where $C_1 = {\mathbb E}_{\bf x} \big( \log p({\bf y}|{\bf x})\big| {\bf y}, {\bm \beta}^{(t)} \big)$ is a term independent of ${\bm \beta}$. Let $Q(\beta_i|{\bm \beta}^{(t)}) = {\mathbb E}_{x_i} \big( \log p(x_i ,\beta_i)| {\bf y}, {\bm \beta}^{(t)} \big)$. We have
\begin{equation}
	Q({\bm \beta} | {\bm \beta}^{(t)}) = C_1 + \sum_{i=1}^M Q(\beta_i|{\bm \beta}^{(t)})
\end{equation}

Since in our setting, the underlying variables are Gaussian, the density $p(x_i|{\bf y}, {\bm \beta}^{(t)})$ produced by the SPA is also Gaussian, with mean $\mu_i$ and variance $\nu_i$. One can explicitly write out $Q(\beta_i|{\bm \beta}^{(t)})$ as
\begin{align}
	Q(\beta_i|{\bm \beta}^{(t)}) &= {\mathbb E}_{x_i}\bigg(\log p(x_i, \beta_i) \big| {\bf y}, {\bm \beta}^{(t)} \bigg) \nonumber \\
	&= {\mathbb E}_{x_i}\Big(\log \big(\frac{1}{\sqrt{2\pi \beta_i}} \exp(-\frac{x_i^2}{2\beta_i}) \: \frac{1}{\beta_i}\big)  \big) \big| {\bf y}, {\bm \beta}^{(t)}\Big) \nonumber \\
	&= C_2 - \frac{3}{2} \log \beta_i - \frac{1}{2\beta_i} {\mathbb E}_{x_i}(x_i^2 | {\bf y},{\bm \beta}^{(t)}) \nonumber \\
	&= C_2 - \frac{3}{2} \log \beta_i - \frac{1}{2\beta_i} (\mu_i^2 + \nu_i),
\end{align}
where $C_2$ is independent of $\beta_i$.

For the \textbf{M-step}, we find 
$${\bm \beta}^{(t+1)} = \arg \max_{{\bm \beta}} Q({\bm \beta} | {\bm \beta}^{(t)}).$$
Clearly $Q({\bm \beta} | {\bm \beta}^{(t)})$ can be maximized by maximizing each $Q(\beta_i|{\bm \beta}^{(t)})$. Hence we have the simple local update rule
\begin{equation}
\beta_i^{(t+1)} = \arg \max_{\beta_i} Q(\beta_i|{\bm \beta}^{(t)}) = \frac{\mu_i^2 + \nu_i}{3}
\end{equation}

We summarize SuPrEM I in Algorithm \ref{alg:suprem}. The inputs to the algorithm contain a stopping criterion ${\cal T}$ and a message-passing schedule ${\cal S}$. The stopping criterion does not really affect the behavior of the algorithm, and there are a few alternatives for a reasonable criterion, which are discussed in Section \ref{sec:sims}. It turns out the message-passing schedule is rather important for achieving the maximum performance. To this end, we develop a message-passing schedule that attains such good performance and we describe this schedule in detail in Appendix I. For all our simulations, we use this fixed schedule. Simulation results indicate that with this fixed schedule, the algorithm is robust in various different scenarios. 
The overall complexity of SuPrEM is $O(M)$ for a fixed number of iterations. We also note that in the presence of noise, the output of the algorithm will not be exactly sparse and a sparse estimate can be constructed using soft-thresholding techniques such as those described in \cite{Nowak2}.


\begin{algorithm}[!tp]
\caption{SuPrEM Algorithm I}
\label{alg:suprem}

\begin{tabbing}
\= \emph{\textbf{Inputs}}: The observed vector ${\bf r}$, the measurement matrix ${\bf F}$, \\ the noise level $\sigma^2$, a stopping criterion ${\cal T}$, and a message-\\passing schedule ${\cal S}$.  \\
\emph{\textbf{1. Initialization:}} Let ${\beta}_k^{(0)} = |({\bf F}^T {\bf r})_k|^2/d_v^2$. 
Initial outgoing \\messages from variable node $x_k$ is $(0, {\beta}_k^{(0)})$.\\

\emph{\textbf{2. Check Nodes:}} For $i=1,2,\dots,N$\\

$\quad$ Let $\{i_1, i_2, \dots, i_{d_c}\}$ be the indices of the variable nodes\\ connected to the $i^{\textrm{th}}$ check node $r_i$. Let the message coming \\from variable node $x_{i_j}$ to the check node $r_i$ at $t^{\textrm{th}}$ iteration\\ be $(\mu_{i_j}^{(t)}, \nu^{(t)}_{i_j})$ for $j = 1, \dots, d_c$. Then the outgoing message \\from check node $r_i$ to variable node $x_{i_j}$ is \\ $\quad \quad \quad (r_i - \sum_{k = 1, k \neq j}^{d_c} \mu_{i_k}^{(t)}, \sum_{k = 1, k \neq j}^{d_c} \nu_{i_k}^{(t)} + \sigma^2)$.\\
$\quad$ The messages are sent according to the schedule ${\cal S}$.\\
\vspace{0.1 cm}
\emph{\textbf{3. Variable Nodes:}} For $k=1,2,\dots, M$\\

$\quad$ Let $\{k_1, k_2, \dots, k_{d_v}\}$ be the indices of the check nodes \\connected to the $k^{\textrm{th}}$ variable node $x_k$. Let the incoming \\message from the check node $r_{k_j}$ to the variable node $x_k$ at \\the $t^{\textrm{th}}$ iteration be $(\mu_{k_j}^{(t)}, \nu_{k_j}^{(t)})$ for $j = 1, \dots, d_v$. \\
$\quad$ \textbf{a.} \emph{EM update:} Let \\$V_k^{(t)} = \bigg(\sum_{j=1}^{d_v} \frac{1}{\nu_{k_j}^{(t)}} + \frac{1}{\beta_k^{(t-1)}}  \bigg)^{-1}$, $\mu_k^{(t)} = V_k^{(t)} \bigg(\sum_{j=1}^{d_v} \frac{\mu_{k_j}^{(t)}}{\nu_{k_j}^{(t)}} \bigg).$\\
$\quad$ Then the EM update is $\beta_k^{(t)} = \frac{(\mu_k^{(t)})^2 + V_k^{(t)}}{3}$.\\

$\quad$ \textbf{b.} \emph{Message updates:} The outgoing message from variable \\
node $x_k$ to check node $r_{k_i}$ at the $(t+1)^{\textrm{th}}$ iteration is given \\by $(\mu_{k_i}^{(t+1)},\nu_{k_i}^{(t+1)}),$ where \\
$\quad \quad \quad \quad \:\:\nu_{k_i}^{(t+1)} = \bigg(\sum_{j=1, j\neq i}^{d_v} \frac{1}{\nu_{k_j}^{(t)}} + \frac{1}{\beta_k^{(t)}}  \bigg)^{-1}$ \\and \\$\quad \quad \quad \quad \mu_{k_i}^{(t+1)} = \nu_{k_i}^{(t+1)} \bigg(\sum_{j=1, j \neq i}^{d_v} \frac{\mu_{k_j}^{(t)}}{\nu_{k_j}^{(t)}}   \bigg).$\\
$\quad$ The messages are sent according to the schedule ${\cal S}$.\\
\emph{\textbf{4. Iterations:}}Repeat (2) and (3) until stopping criterion ${\cal T}$ is \\reached.\\
\emph{\textbf{5. Decisions:}} For the $k^{\textrm{th}}$ variable node $x_k$, let the incoming\\ messages be $(\mu_{k_j}^{({\cal T})}, \nu_{k_j}^{({\cal T})})$ for $j = 1, \dots, d_v$. Let\\
$\quad \quad \quad \quad \quad \hat{V}_k = \bigg(\sum_{j=1}^{d_v} \frac{1}{\nu_{k_j}^{({\cal T})}} + \frac{1}{\beta_k^{({\cal T})}} \bigg)^{-1}$\\
and\\
$\quad \quad \quad \quad \quad \hat{x}_k = \hat{V}_k \bigg(\sum_{j=1}^{d_v} \frac{\mu_{k_j}^{({\cal T})}}{\nu_{k_j}^{({\cal T})}}  \bigg)$.\\
\emph{\textbf{Output:}} The estimate is $\hat{{\bf x}} = (\hat{x}_1, \hat{x}_2, \dots, \hat{x}_M)^T$.
\end{tabbing}
\end{algorithm}


\subsection{SuPrEM Algorithm II}

When the ratio $L/N$ is relatively large, SuPrEM I does not perform well, in particular for high SNRs, since it does not enforce strict sparsity. Thus we propose SuPrEM Algorithm II that enforces sparsity at various stages of the algorithm and sends messages between the nodes of the underlying graph accordingly. To this end, we keep a set of candidate variable nodes ${\cal O}$ that are likely to have non-zero values, and modify the messages from the variable nodes that do not belong to a specified subset of ${\cal O}$ denoted by ${\cal O}_1$. Similar ideas have been used in developing state-of-the-art recovery algorithms for compressive sensing, such as Subspace Pursuit \cite{Dai} and CoSaMP \cite{Needell3}. The full description is given in Algorithm \ref{alg:suprem2}.


\begin{algorithm}[!tp]
\caption{SuPrEM Algorithm II}
\label{alg:suprem2}

\begin{tabbing}
\= \emph{\textbf{Inputs}}: The observed vector ${\bf r}$, the measurement matrix ${\bf F}$, \\ the sparsity level $L$, a stopping criterion ${\cal T}$, the noise level\\ $\sigma^2$ (optional), and a message-passing schedule ${\cal S}$.  \\
\emph{\textbf{1. Initialization:}} Let ${\beta}_k^{(0)} = |({\bf F}^T {\bf r})_k|^2/d_v^2$ and let ${\cal O}_1 = \emptyset$.\\ 
Initial outgoing messages from variable node $x_k$ is $(0, {\beta}_k^{(0)})$.\\

\emph{\textbf{2. Check Nodes:}} Same as in Algorithm I.\\
\vspace{0.1 cm}
\hspace{-0.1cm}\emph{\textbf{3. Variable Nodes:}} Same as in Algorithm I.\\

\emph{\textbf{4. Sparsification:}} \\ \textbf{a.} After the $\beta_k$s have been updated, find the indices of the $L$\\ largest $\beta_k$s. Let these indices be ${\cal O}_2$.\\ \textbf{b.} Merge ${\cal O}_1$ and ${\cal O}_2$, i.e. Let ${\cal O} = {\cal O}_1 \cup {\cal O}_2$.\\
\textbf{c.} For all indices in $k \in {\cal O}$ make a decision on $\hat{x}_k$ (as in Step\\ 5 of Algorithm I). For all indices $k \notin {\cal O}$, let $\hat{x}_k = 0$.\\
\textbf{d.} Identify the indices corresponding to the $L$ largest (in\\ absolute value) coefficients of $\hat{\bf x}$.
Update ${\cal O}_1$ to be this set of\\ $L$ indices.\\
\textbf{e.} The variable vertices $k \in {\cal O}_1$, send out their messages as\\ was decided in Step 3 of Algorithm I. 
The variable vertices\\ $k \notin {\cal O}_1$, send out their messages with $0$ mean and the \\variance that was decided in Step 3 of Algorithm 1.\\

\emph{\textbf{5. Decisions:}} Make decisions only the vertices in ${\cal O}$. Once\\ these are calculated, keep the $L$ indices with the largest\\ $|\hat{x}_k|, k \in {\cal O}$. Set all other indices to 0. \\ 
\emph{\textbf{6. Iterations:}} Repeat (2), (3), (4) and (5) until stopping\\ criterion ${\cal T}$ is reached.\\
\emph{\textbf{Output:}} The estimate is $\hat{{\bf x}} = (\hat{x}_1, \hat{x}_2, \dots, \hat{x}_M)^T$.
\end{tabbing}
\end{algorithm}


The main modification to SuPrEM I is the addition of a sparsification step. Intuitively $\beta_k^{(t)}$ is the reliability of the hypothesis $\hat{x}_k^{(t)} \neq 0$. Throughout the algorithm we maintain a list of variable nodes ${\cal O}_1$ that correspond to the largest $L$ coefficients of $\hat{\bf x}^{(t)}$ at iteration $t$. We also keep a list of variable nodes ${\cal O}_2$ corresponding to the $L$ largest elements of ${\bm \beta}^{(t)}$, i.e. those with the largest reliabilities of the hypothesis $\hat{x}_k^{(t)} \neq 0$. In the sparsification stage, these two sets are merged, ${\cal O} = {\cal O}_1 \cup {\cal O}_2$. The addition and deletion of elements from ${\cal O}$ allow refinements to be made with each iteration. We note $L \leq |{\cal O}| \leq 2L$ at any given iteration. Decisions are made on the elements of ${\cal O}$, and ${\cal O}_1$ is updated. Finally for variable nodes not in ${\cal O}_1$, the mean value of the messages is forced to be 0, but the variance (i.e. the uncertainty about the estimate itself) is kept. By modifying the messages this way, we not only enforce sparsity at the final stage, but also throughout the algorithm.

We note that the noise level $\sigma^2$ is an optional input to the algorithm. Our simulations indicate that the algorithm works without this knowledge also. However, if this extra statistical information is available, it is easily incorporated into the algorithm in a natural way and results in a performance increase.

SuPrEM II has complexity $O(M)$. The only significant operation different than those in SuPrEM I is the determination of the largest $L$ elements of ${\bm \beta}$ and $\hat{\bf x}$. This could be done with $O(M)$ complexity, as described in \cite{whitebook} (Chapter 9). A more straightforward implementation for this stage might use sorting of the relevant coefficients, which would result in a higher complexity of $O(M\log M)$ for the overall algorithm.

\subsection{Reweighted Algorithms} \label{reweigh}

For high $L/N$ ratios, simulation results show that SuPrEM I and SuPrEM II still perform well. However more iterations are needed to achieve very low distortion levels, which may be undesirable. Thus we propose a modification to SuPrEM I and SuPrEM II to speed up the convergence that uses estimates generated within a few iterations. In compressive sensing, employing prior estimates to improve the final solution has been used for $\ell_1$ approximation \cite{Candes-Wakin}, but this increases the running time by a factor of reweighing steps. 

Next, we motivate for our reweighing approach. In our algorithms, the initial choice of $\{ \beta_k^{(0)} = |({\bf F}^T {\bf r})_k|^2/d_v^2 \}$ is based on the intuition that $\beta_k$ must be proportional to $|x_k|^2$. By providing a better estimate for the initial $\{\beta_k^{(0)}\}$, the rate of convergence may be improved. The algorithm is initiated with ${\bm \beta}^{(0)}$ as above and is run for $T_{r_1}$ iterations. 
At the end of this stage, we re-initialize ${\bm \beta}^{(0)'}$ to be
$$\beta_k^{(0)'} = \big|\hat{x}_k^{(T_{r_1})}\big|^2 + \big|({\bf F}^T ({\bf r - F\hat{x}}^{(T_{r_1})}))_k\big|^2/d_v^2,$$
and the algorithm is run for $T_{r_2}$ iterations. This process is repeated recursively until convergence or ${\cal R}$ times. We note that $\sum_{k=1}^{\cal R} T_{r_k} = T$, where $T$ is the original number of fixed iterations. Thus the total number of iterations remains unchanged when we use reweighing.


\section{Simulation Details} \label{sec:sims}

\subsection{Simulation Setup}

In our simulations we used LDFs with parameters $(3,6)$, $(3,12)$ and $(3, 24)$ for $M/N = 2, 4, 8$ and $M = 10000$. We constructed these frames using the progressive edge growth algorithm \cite{peg}, avoiding cycles of length 4 when possible \footnote{We also tested LDFs with 4 cycles and this does not seem to have an adverse effect on the average distortion in the presence of noise.}. Simulations will be presented for SNR$= 12,24,36$ dB, as well as the noiseless case. For various choices of $L$ and SNR, we ran 1000 Monte-Carlo simulations for each value, where ${\bf x}$ is generated as a signal with $L$ non-zero elements that are picked from a Gaussian distribution. The support of ${\bf x}$ is picked uniformly at random. Once ${\bf x}$ is generated, it is normalized such that $||{\bf Fx}||_2 = \sqrt{N}$. Thus SNR$=10 \log_{10}\frac{1}{\sigma^2}$.

Let ${\cal G}$ be the genie decoder that has full information about supp$({\bf x}) = \{i: x_i \neq 0\}$. Let the output of this decoder be $\hat{\bf x}_{genie} = {\cal G}({\bf r})$ obtained by solving the least squares problem involving ${\bf r}$ and the matrix formed by the columns of ${\bf F}$ specified by supp$({\bf x})$. We define the following genie distortion measure: $$\bar{d}_g ({\bf x},\hat{\bf x}_{genie}) = \frac{||{\bf x} - \hat{\bf x}_{genie}||_2^2}{||{\bf x}||_2^2}.$$ This distortion measure is invariant to the scaling of ${\bf x}$ for a fixed SNR. 
For any other recovery algorithm that outputs an estimate $\hat{\bf x}$, we let $$\bar{d}_e ({\bf x},\hat{\bf x}_{e})  = \frac{||{\bf x} - \hat{\bf x}_{e}||_2^2}{||{\bf x}||_2^2},$$ where the subscript $e$ denotes the estimation procedure. We will be interested in the performance of an estimation procedure with respect to the genie decoder. To this end, we define $${\cal D}_{e / g} ({\bf x},\hat{\bf x}_{e}, \hat{\bf x}_{genie}) = \frac{||{\bf x} - \hat{\bf x}_{e}||_2^2}{||{\bf x} - \hat{\bf x}_{genie}||_2^2} = \frac{\bar{d}_e({\bf x},\hat{\bf x}_{e})}{\bar{d}_g({\bf x},\hat{\bf x}_{genie})}.$$ We will be interested in this quantity averaged over $K$ Monte-Carlo simulations, and converted to dB. The closer this quantity is to 0 dB means the closer the performance of the estimation procedure is to the performance of the genie decoder.

In other cases, such as the noiseless case, we will be interested in the empirical probability of recovery. For $K$ Monte-Carlo simulations, this is given by 
$$P_{rec} = \frac1K \sum_{k=1}^K {\mathbb I}({\bf x} \sim \hat{\bf x}_e),$$ where ${\mathbb I}(\cdot)$ is the indicator function for $(\cdot)$ (1 if $(\cdot)$ is true, 0 otherwise). We will define the relation ${\bf x} \sim \hat{\bf x}_e$ to be true only if supp$({\bf x}) = $ supp$(\hat{\bf x}_e)$, unless otherwise specified.

A number of different stopping criterion can be used for ${\cal T}$: 1) $\hat{{\bf x}}$ converges, 2) The minimum value of $\{||{\bf r - F\hat{x}}^{(t)}||_2\}_t$ does not change for $T^d$ iterations , 3) A fixed number of iterations $T$ is reached. In our simulations we use criterion two with $T^d = 30$ and $T = 500$. These values were chosen to make sure that the algorithms did not stop too prematurely. The message passing schedule ${\cal S}$ is described in detail in Appendix \ref{schedules}. Finally, for the reweighted algorithm we use 10 reweighings with $T_{r_1} = \dots = T_{r_{10}} = T/10$.

\subsection{Simulation Results} 

\begin{figure*}[!tp]
\centering
\includegraphics[width=7.2in]{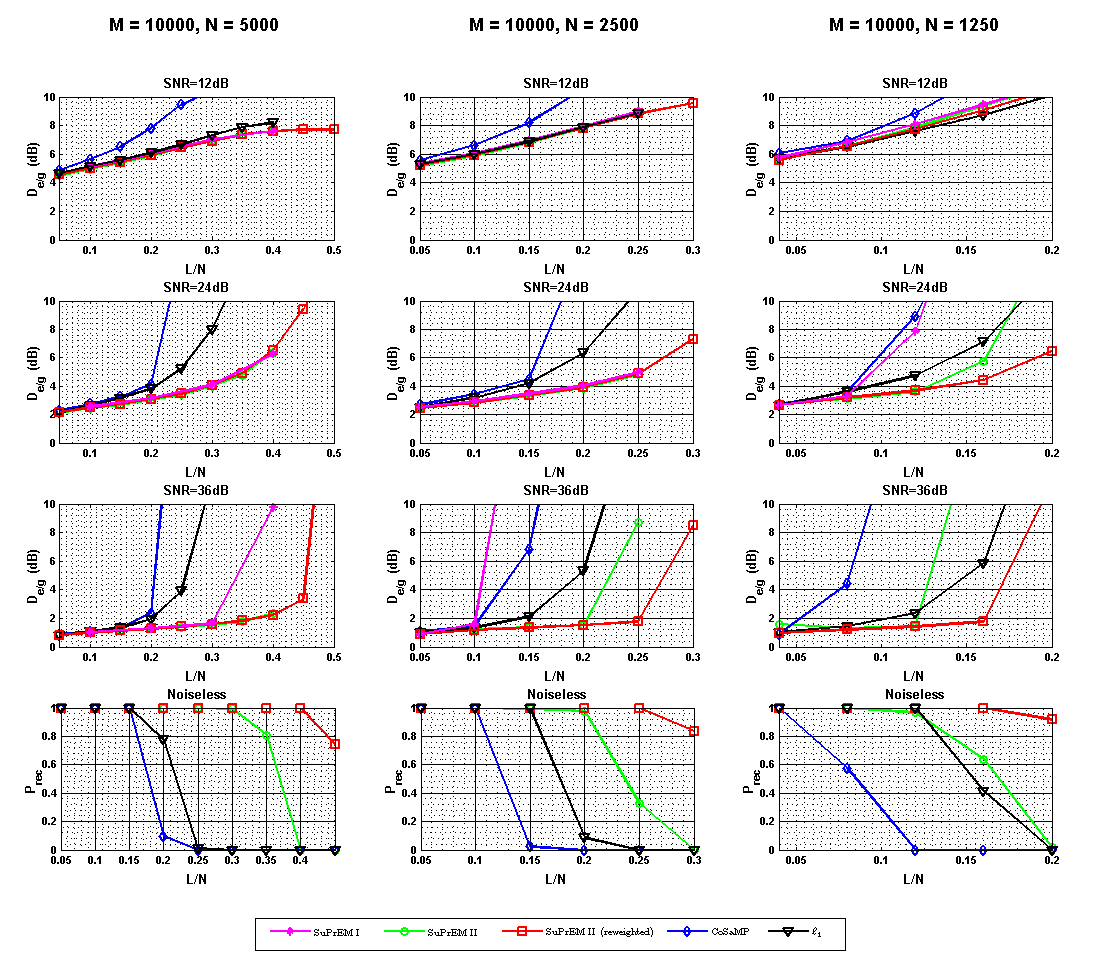}
\caption{Performance comparison of recovery algorithms for sparse signals with Gaussian non-zero components.}
\label{gaussian_results}
\vspace{-.2cm}
\end{figure*}

Simulation results are presented in Figure \ref{gaussian_results} for exactly sparse signals. 

For comparison to our algorithms, we include results for CoSaMP \cite{Needell3} and $\ell_1$ based methods \cite{Candes-Romberg2, Candes-Romberg-Tao, Candes-Tao, Donoho2, gpsr}. For these algorithms we used partial Fourier matrices as measurement matrices. The choice of these matrices is based on their small storage requirements (in comparison to Gaussian matrices), while still satisfying restricted isometry principles. For CoSaMP, we used 100 iterations of the algorithm (and 150 iterations of Richardson's iteration for calculating least squares solutions). For $\ell_1$ based methods, we used the {\fontfamily{pcr}\selectfont L1MAGIC} package in the noiseless case. In the noisy case, we used both {\fontfamily{pcr}\selectfont L1MAGIC}, and the {\fontfamily{pcr}\selectfont GPSR} package (with Barzilai-Borwein Gradient Projection with continuation and debiasing). Since these two methods approximately perform the same, we include the results for {\fontfamily{pcr}\selectfont GPSR} here. In the implementation of GPSR we fine-tune the value of $\tau$ and observe that $\tau = 0.001 ||{\bf F}^T{\bf r}||_{\infty}$ gives the best performance.

Since the outputs of $\ell_1$ based methods and SuPrEM I are not sparse, we threshold ${\bf x}$ to its $L$ largest coefficients and postulate these are the locations of the sparse coefficients. For all methods, we solve the least squares problem involving ${\bf r}$ and the matrix formed by the columns of ${\bf F}$ specified by the final estimate for the locations of the sparse coefficients. For partial Fourier matrices we use Richardson's iteration to calculate this vector, whereas for LDFs we use the LSQR algorithm which also has $O(M)$ complexity \cite{lsqr}. 
 
\subsection{Discussion of The Results} 

The simulation results indicate that the SuPrEM algorithms outperform the other state-of-the-art algorithms. In the low SNR regime (SNR = 12 dB), SuPrEM algorithms and the $\ell_1$ methods have similar performance. In moderate and high SNR regimes, we see that SuPrEM algorithms significantly outperform the other algorithms both in terms of distortion and in terms of the maximum sparsity they can work at. Furthermore for different values of $N$, the maximum sparsity scales as $L = O(N/ \log(M/N))$, which is the same scaling as those of other methods. As we discussed previously the performance of SuPrEM I degrades as sparsity and SNR increases. We also observe that the reweighted SuPrEM II algorithm outperforms the regular SuPrEM II algorithm, even though the maximum number of iterations are the same. 

Finally, compared to the other methods for the noiseless problem, the SuPrEM algorithms can recover signals that have a higher number of non-zero elements. In this case, the reweighted algorithm performs the best, and converges faster. We also note that the results presented for CoSaMP and $\ell_1$ based methods for the noiseless case are optimistic, since we declare success in recovery if $\bar{d}_e ({\bf x},\hat{\bf x}_{e}) <10^{-6}$. We needed to introduce this measure, since these algorithms tend to miss a small portion of the support of ${\bf x}$ containing elements of small magnitude.

We also note that for both partial Fourier matrices and LDFs, the quantity $\bar{d}_g ({\bf x},\hat{\bf x}_{genie})$ is almost the same for a fixed $L$ and SNR. This means that ${\cal D}_{e / g} ({\bf x},\hat{\bf x}_{e}, \hat{\bf x}_{genie})$ provides an objective performance criterion in terms of relative mean-square error with respect to the genie bound, as well as in terms of absolute distortion error $\bar{d}_e ({\bf x},\hat{\bf x}_{e})$.

\subsection{Simulation Results for Natural Images}
For the testing of compressible signals, instead of using artificially generated signals, we used real-world compressible signals. In particular, we compressively sensed the {\fontfamily{pcr}\selectfont db2} wavelet coefficients of the $256 \times 256$ (raw) peppers image using $N = 17000$ measurements. Then we used various recovery algorithms to recover the wavelet coefficients, and we did the inverse wavelet transform to recover the original image. 
\begin{figure*}
\centering
\includegraphics[width=7.2in]{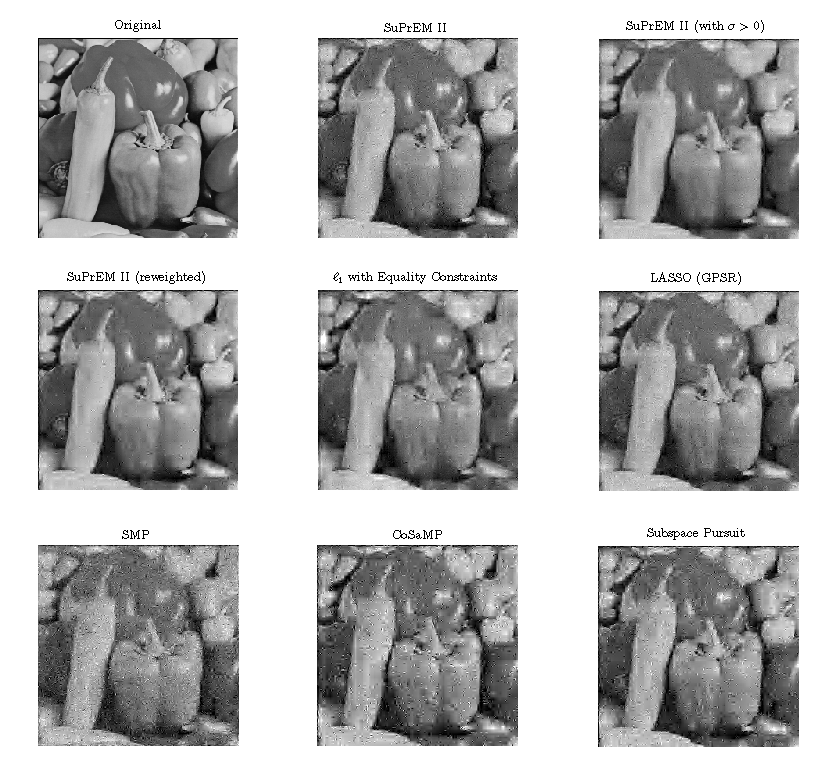}
\caption{Performance comparison of recovery algorithms with a $256 \times 256$ natural image whose {\fontfamily{pcr}\selectfont db2} wavelet coefficients are compressively sensed with N = 17000 measurements.}
\label{pepper_comparison}
\vspace{-.2cm}
\end{figure*}

For SuPrEM algorithms, we used a rate $(3, 12)$ LDF with $M = 68000$ (the wavelet coefficients vector was padded with zeros to match the dimension). We set $L = 8000$ (the maximum sparsity the algorithm converged at) for SuPrEM II. We ran the algorithm first with $\sigma = 0$. We also accomodated for noise, and estimated the per measurement noise to be $\sigma = 0.1\frac{||{\bf r}||_2}{\sqrt{N}}$ and ran the algorithm again\footnote{With this value of $\sigma$, SuPrEM I also provides a similar performance. However since the output in this case is very similar to that of SuPrEM II, we do not include it in the figure.}. We ran our algorithms for just 50 iterations. For the reweighted SuPrEM II algorithm, we let $\sigma = 0$ and we reweighed after 5 steps of the algorithm for a total of 10 reweighings. For SMP, we used the {\fontfamily{pcr}\selectfont SMP} package \cite{Berinde3}. We used a matrix generated by this package, and $L = 8000$. For the remaining methods, we used partial Fourier matrices whose rows were chosen randomly. For $\ell_1$ with equality constraints, we used the {\fontfamily{pcr}\selectfont L1MAGIC} package. For LASSO, we used the {\fontfamily{pcr}\selectfont GPSR} package and $\tau = 0.001 ||{\bf F}^T{\bf r}||_{\infty}$, as described previously, and we thresholded the output to $L=8000$ sparse coefficients and solved the appropriate least squares problem to get the final estimate. For CoSaMP and Subspace Pursuit, we used 100 iterations of the algorithm (and 150 iterations for the Richardson's iteration for calculating the least square solutions). For these algorithms, we used $L = 3000$ for CoSaMP, and $L = 3500$ for Subspace Pursuit. These are slightly lower than the maximum sparsities they converged at ($L = 3500$ and $L=4000$ respectively), but the values we used resulted in better visual quality and PSNR values. 
The results are depicted in Figure \ref{pepper_comparison}.

The PSNR values for the methods are as follows: 23.41 dB for SuPrEM II, 23.83 dB for SuPrEM II (with non-zero $\sigma^2$), 24.79 for SuPrEM II (reweighted), 20.18 dB for CoSaMP, 19.51 dB for SMP, 21.62 dB for $\ell_1$, 23.61 dB for LASSO, 21.27 dB for Subspace Pursuit. 
Among the algorithms that assume no knowledge of noise, we see that SuPrEM II outperforms the other algorithms both in terms of PSNR value and in terms of visual quality. The two algorithms that accomodate noise, SuPrEM II (in this case SuPrEM I also produces a similar output) and LASSO have similar PSNR values. Finally, the reweighted SuPrEM II also assumes no knowledge of noise, and outperforms all other methods by about 1 dB and also in terms of visual quality, without requiring more running time.

\subsection{Further Results}

We studied the effect of the change of degree distributions. For a given $M/N$ ratio, we need to keep the ratio of $d_c/d_v$ fixed however the values can be varied. Thus we compared the performance of $d_v = 3$ LDFs to $d_v = 5$ LDFs, and observed that the latter actually performed sligthly better. However, having a higher $d_v$ means more operations are required. We also observed that the number of iterations required for convergence was slightly higher. Thus we chose to use $d_v = 3$ LDFs that allowed faster decoding. We also note that increasing $d_v$ too much (while keeping $M/N$ fixed) results in performance deterioration, since the graph becomes less sparse, and we run into shorter cycles which affect the performance of SPA. 

We also tested the performance of our constructions and algorithms at $M = 100000$. With $L/M$ and $N/M$ fixed, interestingly the performance improves as $M \to \infty$ for Gaussian sparse signals for a fixed maximum number of 500 iterations. This is in line with intuitions drawn from Shannon Theory \cite{AkTar2}. Another interesting observation is that the number of iterations remain unchanged in this setting. In general, we observed that the number of iterations required for convergence is only a function of $L/M$ and does not change with $M$.


\section{Conclusion}	\label{sec:conc}

In this paper, we constructed an ensemble of measurement matrices with small storage requirements. We denoted the members of this ensemble as Low Density Frames (LDF). For these frames, we provided sparse reconstruction algorithms that have $O(M)$ complexity and that are Bayesian in nature. We evaluated the performance of this ensemble of matrices and their decoding algorithms, and compared their performance to other state-of-the-art recovery algorithms and their associated measurement matrices. We observed that in various cases of interest, SuPrEM algorithms with LDFs outperformed the other algorithms with partial Fourier matrices. In particular, for Gaussian sparse signals and Gaussian noise, we are within 2 dB range of the theoretical lower bound in most cases.

There are various interesting research problems in this area. One is to find a deterministic message-passing schedule that performs as well as (or better than) our probabilistic message-passing schedule and that is amenable to analysis. Another open problem is to analyze the performance of the iterative decoding algorithms for the LDFs theoretically, which may in turn lead to useful design tools (like Density Evolution \cite{Urbanke}) that might help with the construction of LDFs with irregular degree distributions. Adaptive measurements using the soft information available about the estimates, as well as online decoding (similar to Raptor Codes \cite{Raptor}) is another open research area. Finally, if further information is available about the statistical properties of a class of signals (such as block-sparse signals or images represented on wavelet trees as in \cite{modelbased}), the decoding algorithms may be changed accordingly to improve performance.


\appendices

\section{Details On The Message-Passing Schedule} \label{schedules}
 
A message-passing schedule determines the order of messages passed between variable and check nodes of a factor graph. Traditionally, with LDPC codes, the so-called ``flooding'' schedule is used. In this schedule, at each iteration, all the variable nodes pass messages to their neighboring check nodes. Subsequently, all the check nodes pass messages to their neighboring variable nodes. For a cycle-free graph, SPA with a flooding schedule correctly computes a-posteriori probabilities \cite{Bishop, Banihashemi-bunch}. An alternative schedule is the ``serial'' schedule, where we go through each variable node serially and compute the messages to the neighboring nodes. The order in which we go through variable nodes could be lexicographic, random or based on reliabilities. 

In this section, we propose the following schedule based on the intuition derived from our simulations and results from LDPC codes \cite{Banihashemi-prob, Banihashemi-bunch}: For the first iteration, all the check nodes send messages to variable nodes and vice-versa in a flooding schedule. After this iteration, with probability $\frac12$ each check node is ``on'' or ``off''. If a check node is off, it marks the edges connected to itself as an ``inactive'', and sends back the messages it received to the variable nodes. If a check node is on, it marks the edges connected to itself as ``active'' and computes a new message. At the variable nodes, when calculating the new beta, we only use the information coming from active edges. That is for $k=1,2,\dots, M$, let $\{k_1, k_2, \dots, k_{d_v}\}$ be the indices of the check nodes connected to the $k^{\textrm{th}}$ variable node $x_k$. Let the incoming message from the check node $r_{k_j}$ to the variable node $x_k$ at the $t^{\textrm{th}}$ iteration be $(\mu_{k_j}^{(t)}, \nu_{k_j}^{(t)})$ for $j = 1, \dots, d_v$. We will have
 
$$\lambda_k^{(t)} = \bigg(\sum_{(k, k_j) \textrm{ is an active edge}}  \frac{1}{\nu_{k_j}^{(t)}} + \frac{1}{\beta_k^{(t-1)}}  \bigg)^{-1},$$
$$\mu_k^{(t)} = \lambda_k^{(t)} \Bigg(\sum_{(k, k_j)  \textrm{ is an active edge}} \frac{\mu_{k_j}^{(t)}}{\nu_{k_j}^{(t)}}  \Bigg),$$ and
$$\beta_k^{(t)} = \frac{(\mu_k^{(t)})^2 + \lambda_k^{(t)}}{3}.$$

Thus when there is no active edge, we do not perform a $\beta$ update. For the special case when there is only one active edge $(k, k_j)$, we let $\mu_k^{(t)} = \mu_{k_j}$. This is because the intrinsic information is more valuable, and the estimate on $\beta_k^{(t-1)}$ tends to be not as reliable. When we calculate the point estimate, we use all the information at the node, including the reliable and unreliable edges, i.e.
$$\hat{V}_k^{(t)} = \bigg(\sum_{j=1}^{d_v} \frac{1}{\nu_{k_j}^{(t)}} + \frac{1}{\beta_k^{(t)}} \bigg)^{-1},$$
$$\hat{x}_k^{(t)} = \hat{V}_k^{(t)} \bigg(\sum_{j=1}^{d_v} \frac{\mu_{k_j}^{(t)}}{\nu_{k_j}^{(t)}}  \bigg).$$

It is noteworthy that the flooding schedule and serial schedules tend to converge to local minima and they do not perform as well as this schedule we proposed. 


\begin{thebibliography}{99}

\bibitem{AkTar} M. Ak\c{c}akaya and V. Tarokh, ``A Frame Construction and A Universal Distortion Bound for Sparse Representations,'' {\it IEEE Trans. Sig. Proc.}, vol. 56, pp. 2443-2550, June 2008. 
\bibitem{AkTar-isit} M. Ak\c{c}akaya and V. Tarokh, ``On Sparsity, Redundancy and Quality of Frame Representations,'' IEEE Int. Symposium on Information Theory (ISIT), Nice, France, June 2007.
\bibitem{AkTar2} M. Ak\c{c}akaya and V. Tarokh, ``Shannon theoretic limits on noisy compressive sampling,'' arXiv:0711.0366v1 [cs.IT], Nov. 2007.
\bibitem{Andrews} D. Andrews and C. Mallows, ``Scale mixtures of normal distributions,'' {\it J. R. Stat. Soc.}, vol. 36, pp. 99 - 102, 1974.
\bibitem{modelbased} R. Baraniuk, V. Cevher, M. Duarte, and C. Hegde, ``Model-based compressive sensing,'' arXiv:0808.3572v2, Sept. 2008.
\bibitem{Berinde2} R. Berinde, A. C. Gilbert, P. Indyk, H. Karloff, and M. J. Strauss, ``Combining geometry and combinatorics: A unified approach to sparse signal recovery,'' preprint, 2008.
\bibitem{Berinde3} R. Berinde, P. Indyk, and M. Ru\~zi\'c, ``Practical near-optimal sparse recovery in the ell-1 norm,'' Proc. Allerton Conference on Communication, Control, and Computing, Monticello, IL, September 2008.
\bibitem{Dias} J. M. Bioucas-Dias, ``Bayesian Wavelet-Based Image Deconvolution: A GEM Algorithm Exploiting a Class of Heavy-Tailed Priors,'' {\it IEEE Trans. Image Proc.}, vol. 15, pp. 937-951, Apr. 2006.
\bibitem{Bishop} C. M. Bishop, {\it Pattern Recognition and Machine Learning}, First Edition, Springer, New York, NY, 2006.
\bibitem{Candes-Romberg2} E. J. Cand\`es, J. Romberg, ``Practical signal recovery from random projections,'' presented at the {\it Wavelet Appl. Signal Image Process. XI, SPIE Conf.}, San Diego, CA, 2005.
\bibitem{Candes-Romberg-Tao} E. J. Cand\`es, J. Romberg, T. Tao, ``Stable signal recovery for incomplete and inaccurate measurements,'' {\it Commun. Pure Appl. Math.}, vol. 59, pp. 1207-1223, Aug. 2006.
\bibitem{Dantzig} E. J. Cand\`es and T. Tao, ``The Dantzig selector: statistical estimation when p is much larger than n,'' Annals of Statistics, 35, pp. 2313-2351, Dec. 2007.
\bibitem{Candes-Tao} E. J. Cand\`es, T. Tao, ``Decoding by Linear Programming,'' {\it IEEE Trans. Inf. Theory}, vol. 51, pp. 4203-4215, Dec. 2005.
\bibitem{Candes-Wakin} E. J. Cand\`es, M. Wakin and S. Boyd, ``Enhancing sparsity by reweighted l1 minimization,'' J. Fourier Anal. Appl., vol. 14, pp. 877-905.
\bibitem{whitebook} T. Cormen, C. Lesierson, L. Rivest, and C. Stein, {\it Introduction to Algorithms}, Second Edition, MIT Press, Cambridge, MA, 2001.
\bibitem{Donoho2} D. L. Donoho, ``Compressed Sensing,'' {\it IEEE Trans. Inf. Theory}, vol. 52, pp. 1289-1306, April 2006.
\bibitem{Dai} W. Dai and O. Milenkovic, ``Subspace pursuit for compressive sensing: Closing the gap between performance and complexity,'' arXiv:0803.0811v2 [cs.NA], March 2008. 
\bibitem{Nowak2} M. A. T. Figueiredo and R. Nowak, ``Wavelet-based image estimation: An empirical bayes approach using Jeffreys' noninformative prior,'' IEEE Trans. Image Proc., vol. 10, pp. 1322-1331, Sep. 2001.
\bibitem{gpsr} M. A. T. Figueiredo, R. D. Nowak and S. J. Wright, ``Gradient projection for sparse reconstruction: Application to compressed sensing and other inverse problems,'' \emph{IEEE Journal of Selected Topics in Signal Processing}, vol. 1, pp. 586-598, Dec. 2007.
\bibitem{Goyal1} A. K. Fletcher, S. Rangan and V. K. Goyal, ``Necessary and Sufficient Conditions on Sparsity Pattern Recovery,'' arXiv:0804.1839v1 [cs.IT], Apr. 2008.
\bibitem{Gallager} R. G. Gallager, {\it Low-Density Parity-Check Codes}, MIT Press, Cambridge, MA, 1963.
\bibitem{peg} X.-Y. Hu, E. Eleftheriou and D. M. Arnold, ``Regular and irregular progressive edge-growth tanner graphs,'' {\it IEEE Trans. Inf. Theory}, vol. 51, pp. 386-398, Jan. 2005.
\bibitem{Carin} S. Ji, Y. Xue and L. Carin, ``Bayesian compressive sensing,'' {\it IEEE Trans. on Sig. Proc.}, vol. 56, pp. 2346-2356, June 2008.
\bibitem{FactorGraphs} F. R. Kschischang, B. J. Frey, and H.-A. Loeliger, ``Factor Graphs and the Sum-Product Algorithm,'' {\it IEEE Trans. Inf. Theory}, vol. 47, pp. 498-519, Feb. 2001.
\bibitem{Mackay} D. J. C. MacKay, ``Good error correcting codes based on very sparse matrices,'' {\it IEEE Trans. Inf. Theory}, vol. 45, pp. 399-431, Mar. 1999.
\bibitem{Mackay-book} D. J. C. MacKay, {\it Information Theory, Inference, and Learning Algorithms}, First Edition, Cambridge University Press, Cambridge, UK, 2002.
\bibitem{Banihashemi-prob} Y. Mao and A. H. Banihashemi, ``Decoding Low-Density Parity-Check Codes With Probabilistic Scheduling,'' {\it IEEE Comm. Letters}, vol. 5, pp. 414-416, Oct. 2001.
\bibitem{Krishnan} G. J. McLachlan and T. Krishnan, {\it The EM Algorithm and Extensions}, First Edition, John Wiley \& Sons, New York, NY, 1997.
\bibitem{Moon} T. K. Moon, ``The EM algorithm in signal processing,'' {\it IEEE Sig. Proc. Mag.}, vol. 13, pp. 47-60, Nov. 1996.
\bibitem{Needell3} D. Needell and J. A. Tropp, ``CoSaMP: Iterative signal recovery from incomplete and inaccurate samples,'' arXiv:0803.2392v2 [math.NA], Apr. 2008.
\bibitem{lsqr} C. C. Paige and M. A. Saunders, ``LSQR: Sparse Linear Equations and Least Squares Problems,'' \emph{ACM Transactions on Mathematical Software (TOMS)}, vol. 8, pp.195-209, June 1982.
\bibitem{Portilla} J. Portilla, V. Strela, M. J. Wainwright and E. P. Simoncelli, ``Image Denoising Using Scale Mixtures of Gaussians in the Wavelet Domain,'' {\it IEEE Trans. Image Proc.}, vol. 12, pp. 1338-1351, Nov. 2003.
\bibitem{Urbanke} T. J. Richardson and R.L. Urbanke, ``The capacity of low-density parity-check codes under message passing decoding,'' {\it IEEE Trans. Inf. Theory}, vol. 47, no. 2, pp. 599-618, Feb. 2001.
\bibitem{Robert} C. Robert, {\it The Bayesian Choice: A Decision Theoretic Motivation}, First Edition, New York, NY, Springer-Verlag, 1994.
\bibitem{Baron1} S. Sarvotham, D. Baron, and R. Baraniuk, ``Sudocodes - Fast Measurement and Reconstruction of Sparse Signals,'' {\it Proc. IEEE Int. Symp. on Inf. Theory (ISIT)}, Seattle, WA, July 2006.
\bibitem{Baron2} S. Sarvotham, D. Baron, and R. Baraniuk, ``Compressed Sensing Reconstruction via Belief Propagation,'' preprint, 2006.
\bibitem{Raptor} A. Shokrollahi, ``Raptor codes,'' {\it IEEE Trans. Inf. Theory}, vol. 52, pp. 2551-2567, June 2006.
\bibitem{Sipser} M. Sipser and D. A. Spielman,``Expander codes,'' {\it IEEE Trans. Inf. Theory}, vol. 42, pp. 1710-1722, Nov. 1996.
\bibitem{Tanner} R. M. Tanner, `` A Recursive Approach to Low Complexity Codes,'' {\it IEEE Trans. Inf. Theory}, vol. 27, pp. 533-547, Sept. 1981.
\bibitem{Tipping} M. E. Tipping, ``Sparse Bayesian learning and the relevance vector machine,'' Journal of Machine Learning Research, vol. 1, pp. 211-244, 2001.
\bibitem{Tipping2} M. E. Tipping, ``Bayesian inference: An introduction to principles and practice in machine learning,'' in O. Bousquet, U. von Luxburg, and G. Rätsch (Eds.), Advanced Lectures on Machine Learning, pp. 41-62, Springer, 2004.
\bibitem{Tropp} J. A. Tropp, ``Topics in Sparse Approximation'', Ph.D. dissertation, Computational and Applied Mathematics, UT-Austin, August 2004.
\bibitem{Tropp2} J. A. Tropp, ``Just relax: Convex programming methods for identifying sparse signals'', {\it IEEE Trans. Inf. Theory}, vol. 51, no. 3, pp. 1030-1051, Mar. 2006.
\bibitem{Gilbert} J. A. Tropp, A. C. Gilbert, ``Signal recovery from partial information via Orthogonal Matching Pursuit'', {\it IEEE Trans. Inf. Theory}, vol. 53, pp.4655-4666, Dec. 2007.
\bibitem{Wainwright} M. J. Wainwright, ``Information-Theoretic Limits on Sparsity Recovery in the High-Dimensional and Noisy Setting,'' Technical Report, UC Berkeley, Department of Statistics, Jan. 2007.
\bibitem{Wainwright2} M. J. Wainwright, ``Sharp thresholds for noisy and high-dimensional recovery of sparsity using ${\ell}_1$-constrained quadratic programming,'' Technical report, UC Berkeley, Department of Statistics, May 2006. 
\bibitem{Weiss} Y. Weiss and W. T. Freeman, ``Correctness of belief propagation in Gaussian graphical models of arbitrary topology,'' {\it Proc. Adv. Neural Inform. Processing Syst.}, vol. 12, Dec. 1999.
\bibitem{Wiberg} N. Wiberg, ``Codes and decoding on general graphs,'' Ph.D. dissertation, Link\"oping University, Sweden, 1996.
\bibitem{Banihashemi-bunch} H. Xiao and A. H. Banihashemi, ``Graph-Based Message-Passing Schedules for Decoding LDPC Codes,'' {\it IEEE Trans. on Comm.}, vol. 52, pp. 2098-2105, Dec. 2004.
\bibitem{Xu} W. Xu and B. Hassibi, ``Efficient compressive sensing with deterministic guarantees using expander graphs,'' {\it Proc. IEEE Inf. Theory Workshop}, Lake Tahoe, CA, Sept. 2007.


\end{thebibliography}
\end{document}